\documentstyle[aps,amssymb]{revtex}

\begin{document}
\draft
\title{Experimental demonstration of an interferometric complementarity of one- and
two-particle interference in a bulk nuclear magnetic resonance ensemble}
\author{Xinhua Peng$^{\thanks{%
Corresponding author: E-mail: xhpeng@wipm.ac.cn; {Fax: 0086-27-87199291.}}}$%
, Xiwen Zhu, Maili Liu, and Kelin Gao}
\address{State Key Laboratory of Magnetic Resonance and Atomic and Molecular Physics,%
\\
Wuhan Institute of Physics and Mathematics, The Chinese Academy of Sciences,%
\\
Wuhan, 430071, People's Republic of China}
\maketitle

\begin{abstract}
We analyze an interferometric complementarity between one- and two-particle
interference in the general case: $V_{i}^{2}+V_{12}^{2}\leq 1$ $(i=1$, $2)$,
and further examine the relation among one-particle interference visibility $%
V_{i}$, two-particle interference visibility $V_{12}$ and the predication $%
P_{i}$ of the path of a single particle. An equality $%
V_{i}^{2}+V_{12}^{2}+P_{i}^{2}=1$ $(i=1$, $2)$ is achieved for any pure
two-particle source, which implies the condition of the complementarity
relation to reach the upper bound and its relation to another
interferometric complementarity between path information and interference
visibility of a single particle. Meanwhile, the relationships of the
complementarities and the entanglement $E$ of the composite system are also
investigated. Using nuclear magnetic resonance techniques, the two-particle
interferometric complementarity was experimentally tested with the
ensemble-averaged spin states, including two extreme cases and an
intermediate case.
\end{abstract}


\section{Introduction}

Quantum superposition and its resulting interference are arguably the most
fundamental effects in quantum mechanics, which leads to the concept of
complementarity. In 1927, Bohr\cite{Bohr} first reviewed this subject,
claimed that the wave- and particle-like behaviors of a quantum mechanical
object are mutually exclusive in a single experiment and expressed it as
complementarity. So Bohr complementarity is often superficially identified
with the `wave-particle duality'\cite{Feynman,Scully} which emphasizes two
extreme cases that each experiment must be described in terms of either
waves or particles, as explained in the textbook. Theoretical investigations
about the intermediate situations in which particle and wave aspects coexist
have led to some quantitative statements about wave-particle duality\cite
{Wootters,Bartell,Greenberger,Mandel,JaegerSV,Englert,EnglertB}, some of
which are expressed as inequalities $V^{2}+P^{2}\leq 1$ or $V^{2}+D^{2}\leq 1
$ about the complementarity between `which-way' (WW) information of a
particle: the predictability $P$ and the distinguishability $D$ of the path,
and the visibility $V$ of interference fringes\cite
{JaegerSV,Englert,EnglertB}. Bohr complementarity is often illustrated by
means of interferometers and has been experimentally investigated in
one-particle interferometer with individual particles, including photons\cite
{Taylor,Mittelstaedt,Schwindt}, electrons\cite{Mollenstedt}, neutrons\cite
{Zeilinger,Rauch,Summhammer}, atoms\cite{Carnal,Durr,Bertet} and nuclear
spins in a bulk ensemble with nuclear magnetic resonance (NMR) techniques%
\cite{Zhu,Peng}.

Complementarity, however, is a more general concept which states that
quantum systems possess properties that are equally real but mutually
exclusive. Naturally, the techniques of standard one-particle interferometry
have been extended to two-particle interferometry by employing correlated
two-particle systems. The two-particle interference phenomenon was first
observed by Ghosh and Mandel\cite{Ghosh} using photon pair produced by
parametric down-conversion, where the nonclassical effects associated with
Einstein-Podolsky-Rosen (EPR) states were exhibited. Since that,
two-particle interferometry has been intensively studied theoretically and
experimentally\cite{All}. Whereafter, a fact between one- and two-particle
interference was noticed by Horne and Zeilinger\cite{HorneZeil}, i.e., when
the two-particle visibility is unity, the one-particle visibility is zero,
and vice versa. The visibilities of the one- and two-particle interference
are described as a pair of complementary quantities. A systematic
investigation of intermediate cases was carried out by Jaeger et al.\cite
{JaegerHS}, who showed that in a large family of states $\left| \Theta
\right\rangle $ an interferometric complementarity holds between
one-particle visibility $V_{_{i}}$ and two-particle visibility $V_{_{12}}$: 
\begin{equation}
V_{_{i}}^{2}+V_{_{12}}^{2}\leq 1,(i=1,2)
\end{equation}
where the upper bound is saturated for a special family of states while most
of the cases are limited in the inequality (1). However, the condition of
achieving the upper bound has not yet been pointed out. The complementarity
relation of one- and two-photon interference has been experimentally
demonstrated in a Young's double-slit experiment by Abouraddy et al.\cite
{Abouraddy}. Except this experiment, we have not still found other
experimental versions in the intermediate regime. Moreover, inequality (1)
can be strengthened to an equality with a suitable extension of operations
on the pair of particles\cite{JaegerSV}, but this conclusion is based on the
different parameters of experimental configuration, that is, one can not
observe such interference fringes to obtain $V_{_{i}}$ and $V_{_{12}}$ that
satisfy the equality in practice\cite{JaegerSV}.

Then, what is the condition of this complementarity being of the form of an
equality $V_{_{i}}^{2}+V_{_{12}}^{2}=1$? And what is the reason why the
complementarity relation is restricted in the inequality (1)? In this paper,
we analyze the one- and two-particle interference complementarity in the
general case and examine the relation among $V_{i}$, $V_{12}$ and the
predication of the path $P_{i}$ to obtain an equality: $%
V_{i}^{2}+V_{12}^{2}+P_{i}^{2}=1,\quad (i=1,2)$ for any pure two-particle
source, which gives the answers of the two questions above, and, in fact,
embodies two interferometric complementarity relations whose relations are
discussed. Furthermore, we also clarify the relationships of $V_{12},$ the
distinguishability of the path $D_{i}$ and the entanglement $E$ of the
composite system which show an important view that this complementarity can
arise due to the correlations. Using NMR techniques detailed in our works%
\cite{Zhu,Peng}, the one- and two-particle complementarity relation of $%
V_{i}^{2}+V_{12}^{2}=1$ with $P_{i}=0$ has been experimentally tested with
the ensemble-averaged spin states of $^{13}$C and $^{1}$H nuclei in $^{13}$%
C-labelled chloroform $^{13}$CHCl$_{3}$, including two extreme cases:
perfect two-particle interference fringe and no one-particle interference
fringe for an entangled state and the opposite for a product state, and an
intermediate case of a special family of states, by measuring both single
and joint probability densities of spin states.

\section{Complementarity in two extreme cases}

We briefly review some dinfinitions and two-particle interferometric
complementarity in two extreme cases. A schematic arrangement for a
two-particle interferometer with variable phase shifters is given in Fig. 1%
\cite{JaegerHS}. The source S emits a pair of particles 1+2, one of which
propagates along the path A and/or A$^{\prime }$, through a variable phase
shifter $\phi _{1}$ impinging on an ideal symmetric beam splitter BS$_{1}$,
and is then registered in either beam $K_{1}$ or $L_{1}$, while on the other
side there is the analogous process on the other (see Fig. 1). The locution
``and/or'' refers in a brief way to a quantum-mechanic superposition, the
details of which are specified by the state of particles 1+2. Assuming that
the pair of particles 1+2 are ideally two-level or spin-1/2 quantum
entities, the Hilbert space H$_{1}$ associated with particle 1 is spanned by
vectors $\left| A\right\rangle $ and $\left| A^{\prime }\right\rangle $
representing states of propagation in the paths A and A$^{\prime }$, and the
Hilbert space H$_{2}$ associated with particle 2 by analogous vectors $%
\left| B\right\rangle $ and $\left| B^{\prime }\right\rangle $. In the same
way, the space of the output states is the subspaces spanned by the vectors $%
\left| K_{1}\right\rangle $ and $\left| L_{1}\right\rangle $ for particle 1
and by $\left| K_{2}\right\rangle $ and $\left| L_{2}\right\rangle $ for
particle 2.

According to the standard definition of visibility $V=(I_{max}-I_{min})/%
\left( I_{max}+I_{min}\right) $, the usual definition for single-particle
fringe visibility is 
\begin{equation}
V_{i}=\frac{\left[ p\left( K_{i}\right) \right] _{\max }-\left[ p\left(
K_{i}\right) \right] _{\min }}{\left[ p\left( K_{i}\right) \right] _{\max
}+\left[ p\left( K_{i}\right) \right] _{\min }},\qquad (i=1,2),
\end{equation}
whereas for two-particle fringe visibility the ''corrected'' definition\cite
{JaegerSV,JaegerHS} is used: 
\begin{equation}
V_{12}=\frac{\left[ \overline{p}\left( K_{1}K_{2}\right) \right] _{\max
}-\left[ \overline{p}\left( K_{1}K_{2}\right) \right] _{\min }}{\left[ 
\overline{p}\left( K_{1}K_{2}\right) \right] _{\max }+\left[ \overline{p}%
\left( K_{1}K_{2}\right) \right] _{\min }},
\end{equation}
where the ``corrected'' joint probability $\overline{p}\left(
K_{1}K_{2}\right) =p\left( K_{1}K_{2}\right) -p\left( K_{1}\right) p\left(
K_{2}\right) +\frac{1}{4}$ to assure the reasonableness of the visibility $%
V_{12}$\cite{JaegerSV,JaegerHS}, and $p\left( K_{i}\right) $ and $p\left(
K_{1}K_{2}\right) $ denote the probabilities of single and joint detections,
respectively. These visibilities can be obtained by observing the variations
of both single and joint probabilities as functions of the phases $\phi _{1}$
and $\phi _{2}$. For simplicity, our experiments take the case of $\left|
A\right\rangle =\left| B\right\rangle =\left| K_{1}\right\rangle =\left|
K_{2}\right\rangle =\left| \uparrow \right\rangle =\left( 
\begin{array}{l}
1 \\ 
0
\end{array}
\right) $ and $\left| A^{\prime }\right\rangle =\left| B^{\prime
}\right\rangle =\left| L_{1}\right\rangle =\left| L_{2}\right\rangle =\left|
\downarrow \right\rangle =\left( 
\begin{array}{l}
0 \\ 
1
\end{array}
\right) $ representing the single-particle states with spin-up and
spin-down, respectively.

Consider two extreme cases\cite{JaegerHS}: when particles 1+2 is prepared in
an entangled state 
\begin{equation}
\left| \Psi \right\rangle =\frac{1}{\sqrt{2}}\left( \left| \uparrow
\right\rangle _{1}\left| \uparrow \right\rangle _{2}+\left| \downarrow
\right\rangle _{1}\left| \downarrow \right\rangle _{2}\right) ,
\end{equation}
or in a product state 
\begin{equation}
\left| \Phi \right\rangle =\left[ \frac{1}{\sqrt{2}}\left( \left| \uparrow
\right\rangle _{1}+\left| \downarrow \right\rangle _{1}\right) \right]
\otimes \left[ \frac{1}{\sqrt{2}}\left( \left| \uparrow \right\rangle
_{2}+\left| \downarrow \right\rangle _{2}\right) \right] .
\end{equation}
The joint action of variable phase shifters and symmetric beam splitters can
be described by a unitary operation 
\begin{equation}
U\left( \phi _{1},\phi _{2}\right) =U_{1}\left( \phi _{1}\right) \otimes
U_{2}\left( \phi _{2}\right) 
\end{equation}
where 
\begin{equation}
U_{i}\left( \phi _{i}\right) =\frac{1}{\sqrt{2}}\left( 
\begin{array}{cc}
1 & e^{i\phi _{i}} \\ 
-e^{-i\phi _{i}} & 1
\end{array}
\right) 
\end{equation}
and subscript $i$ represents particle $i$. $U\left( \phi _{1},\phi
_{2}\right) $ is the unitary mapping from the space of input states of the
pair of particles 1+2 into the space of output states. It is easy to show
that the joint and single measurement probabilities in the space of output
states are: for the $\left| \Psi \right\rangle $ state of Eq.(4), 
\begin{equation}
\begin{array}{l}
p\left( \left| \uparrow \right\rangle _{1}\left| \uparrow \right\rangle
_{2}\right) =p\left( \left| \downarrow \right\rangle _{1}\left| \downarrow
\right\rangle _{2}\right) =\frac{1}{4}\left[ 1+\cos \left( \phi _{1}+\phi
_{2}\right) \right]  \\ 
p\left( \left| \uparrow \right\rangle _{1}\left| \downarrow \right\rangle
_{2}\right) =p\left( \left| \downarrow \right\rangle _{1}\left| \uparrow
\right\rangle _{2}\right) =\frac{1}{4}\left[ 1-\cos \left( \phi _{1}+\phi
_{2}\right) \right]  \\ 
p\left( \left| \uparrow \right\rangle _{1}\right) =p\left( \left| \downarrow
\right\rangle _{1}\right) =p\left( \left| \uparrow \right\rangle _{2}\right)
=p\left( \left| \downarrow \right\rangle _{2}\right) =\frac{1}{2}
\end{array}
\end{equation}
and for the $\left| \Phi \right\rangle $ state of Eq.(5), 
\begin{equation}
\begin{array}{l}
p\left( \left| x\right\rangle _{1}\left| y\right\rangle _{2}\right) =p\left(
\left| x\right\rangle _{1}\right) p\left( \left| y\right\rangle _{2}\right) 
\\ 
p\left( \left| \uparrow \right\rangle _{i}\right) =\frac{1}{2}\left( 1+\cos
\phi _{i}\right)  \\ 
p\left( \left| \downarrow \right\rangle _{i}\right) =\frac{1}{2}\left(
1-\cos \phi _{i}\right) 
\end{array}
\end{equation}
where $x,y=\uparrow $or$\downarrow $. Obviously, the probabilities of joint
measurements in Eqs.(8) exhibit cosinusoidal modulations via the variable
phase $\phi _{i}$ which yields two-particle interference fringes for the
entangled state $\left| \Psi \right\rangle $, whereas there is no
one-particle interference fringes. Conversely, a cosinusoidal dependence of
the single probabilities on the variable phase $\phi _{i}$ in Eqs.(9) yields
one-particle interference fringes for the product state $\left| \Phi
\right\rangle $, but no genuine two-particle fringes, in that the variation
of the joint probability $p\left( \left| x\right\rangle _{1}\left|
y\right\rangle _{2}\right) $ results only from the variations of the single
probabilities $p\left( \left| x\right\rangle _{1}\right) $ and $p\left(
\left| y\right\rangle _{2}\right) $. This is the reason for using the
``corrected'' definition for two-particle fringe visibility. Further,
Eqs.(8) shows $p\left( \left| x\right\rangle _{1}\left| y\right\rangle
_{2}\right) \neq p\left( \left| x\right\rangle _{1}\right) p\left( \left|
y\right\rangle _{2}\right) $, a manifestation of quantum nonlocality
associated with the entangled state $\left| \Psi \right\rangle $. From the
definitions of Eqs.(2) and (3), the theoretical outcomes of the visibilities
are found to be $V_{1}=V_{2}=0,V_{12}=1$ for $\left| \Psi \right\rangle $
and $V_{1}=V_{2}=1,V_{12}=0$ for $\left| \Phi \right\rangle $.

The phenomenon can also be partially explained via the spirit of the
`which-way' (WW) interferometry experiment to test Bohr complementarity: any
attempt to gain WW information unavoidably destroys the interference pattern%
\cite{Feynman,Scully}. For example, when the composite system of particles
1+2 is prepared in the entangled state $\left| \Psi \right\rangle $, WW
information of one particle is stored in the states of the other through
their correlation. Consequently, as far as a single particle (1 or 2) is
concerned, there are full WW information and no one-particle fringes.
However, for the pair of particles 1+2, there is no way to determine whether
1+2 takes the composite path $\left| \uparrow \right\rangle _{1}\left|
\uparrow \right\rangle _{2}$ or the composite path $\left| \downarrow
\right\rangle _{1}\left| \downarrow \right\rangle _{2}$, and hence perfect
two-particle interference pattern displays. In contrast, if the system is in
the product state $\left| \Phi \right\rangle ,$ it is impossible to
determine the path of one particle through the other, and thus there is
one-particle interference.

\section{Complementarity in the general case}

Now, we turn to the general case from two extreme $\left| \Psi \right\rangle 
$ and $\left| \Phi \right\rangle $. The most general state of pure
two-particle source 1+2 that can be formed from the basis of $\left\{ \left|
\uparrow \right\rangle _{i},\left| \downarrow \right\rangle _{i}\right\} $
is 
\begin{equation}
\left| \Theta \right\rangle =\gamma _{1}\left| \uparrow \right\rangle
_{1}\left| \uparrow \right\rangle _{2}+\gamma _{2}\left| \uparrow
\right\rangle _{1}\left| \downarrow \right\rangle _{2}+\gamma _{3}\left|
\downarrow \right\rangle _{1}\left| \uparrow \right\rangle _{2}+\gamma
_{4}\left| \downarrow \right\rangle _{1}\left| \downarrow \right\rangle _{2},
\end{equation}
where 
\begin{equation}
\left| \gamma _{1}\right| ^{2}+\left| \gamma _{2}\right| ^{2}+\left| \gamma
_{3}\right| ^{2}+\left| \gamma _{4}\right| ^{2}=1,
\end{equation}
where $\gamma _{i}=\left| \gamma _{i}\right| e^{i\delta _{i}}$ are complex
numbers. As the visibilities are independent of the phases $\delta _{i}$, in
our analysis, all of $\gamma _{i}$ are taken to be real numbers\cite
{JaegerHS}.

Likewise, the unitary operation $U$ in Eq.(7) is applied to $\left| \Theta
\right\rangle $ and straightforward calculations yield 
\begin{equation}
\begin{array}{l}
p\left( \left| \uparrow \right\rangle _{1}\right) =\frac{1}{2}\left[
1+2\left( \gamma _{1}\gamma _{3}+\gamma _{2}\gamma _{4}\right) \cos \phi
_{1}\right] , \\ 
p\left( \left| \downarrow \right\rangle _{1}\right) =\frac{1}{2}\left[
1-2\left( \gamma _{1}\gamma _{3}+\gamma _{2}\gamma _{4}\right) \cos \phi
_{1}\right] , \\ 
p\left( \left| \uparrow \right\rangle _{2}\right) =\frac{1}{2}\left[
1+2\left( \gamma _{1}\gamma _{2}+\gamma _{3}\gamma _{4}\right) \cos \phi
_{2}\right] , \\ 
p\left( \left| \downarrow \right\rangle _{2}\right) =\frac{1}{2}\left[
1-2\left( \gamma _{1}\gamma _{2}+\gamma _{3}\gamma _{4}\right) \cos \phi
_{2}\right] ,
\end{array}
\end{equation}
and 
\begin{equation}
\overline{p}\left( \left| \uparrow \right\rangle _{1}\left| \uparrow
\right\rangle _{2}\right) =\frac{1}{4}\left[ 1+M\cos \phi _{1}\cos \phi
_{2}-N\sin \phi _{1}\sin \phi _{2}\right] 
\end{equation}
where 
\begin{equation}
\begin{array}{l}
N=2\left( \gamma _{1}\gamma _{4}-\gamma _{2}\gamma _{3}\right)  \\ 
M=2\left( \gamma _{1}\gamma _{4}+\gamma _{2}\gamma _{3}\right) -4\left(
\gamma _{1}\gamma _{3}+\gamma _{2}\gamma _{4}\right) \left( \gamma
_{1}\gamma _{2}+\gamma _{3}\gamma _{4}\right) 
\end{array}
,
\end{equation}
and the similar expressions for other $\bar{p}\left( \left| x\right\rangle
_{1}\left| y\right\rangle _{2}\right) $. It is readily verified that $\left|
N\right| \geqslant \left| M\right| $ by using the normalization condition of
Eq.(11). With the derivative method\cite{JaegerHS}, the maximal and minimal
values of $\overline{p}\left( \left| \uparrow \right\rangle _{1}\left|
\uparrow \right\rangle _{2}\right) $ are deduced as: $\overline{p}_{\max
,\min }\left( \left| \uparrow \right\rangle _{1}\left| \uparrow
\right\rangle _{2}\right) =\frac{1}{4}\left( 1\pm \left| N\right| \right) $,
which can be achieved only when $\left( \cos \phi _{1}\cos \phi _{2},\sin
\phi _{1}\sin \phi _{2}\right) =\left( 0,\pm 1\right) $, i.e., $\left( \phi
_{1},\phi _{2}\right) =\left( n\pi +\pi /2,m\pi +\pi /2\right) ,\left(
n,m=0,\pm 1,...\right) $. Hence, on substituting for the maximal and minimal
values of these probabilities in Eqs.(2) and (3), one gets

\begin{equation}
\begin{array}{l}
V_{1}=\left| 2\left( \gamma _{1}\gamma _{3}+\gamma _{2}\gamma _{4}\right)
\right| , \\ 
V_{2}=\left| 2\left( \gamma _{1}\gamma _{2}+\gamma _{3}\gamma _{4}\right)
\right| , \\ 
V_{12}=\left| 2\left( \gamma _{1}\gamma _{4}-\gamma _{2}\gamma _{3}\right)
\right| ,
\end{array}
\end{equation}
and 
\begin{equation}
V_{i}^{2}+V_{12}^{2}=1-\left[ \left( \gamma _{2}^{2}-\gamma _{3}^{2}\right)
-\left( -1\right) ^{i}\left( \gamma _{1}^{2}-\gamma _{4}^{2}\right) \right]
^{2}\leq 1,\quad \left( i=1,2\right) .
\end{equation}
The complementarity relation is achieved for any pure state of a composite
two-particle system. In this type of interference experiments, another
interesting physical quantity is the WW information of one particle when the
other serves as the WW maker. Because the two complementarities --- between
path distinguishability and single-particle visibility and between one- and
two-particle visibilities --- are intimately connected, the one-particle
visibility $V_{i}$ enters in the same way in both of them\cite{JaegerSV}.
Without any measurement, the available {\it a priori} WW knowledge of one
particle is described by the predictability $P_{i}$ of the alternatives,
which represents Man's knowledge before measure; while the
distinguishability $D_{i}$ denotes the maximal information about the
``path'' that can be extracted from an appropriate measure, which is
Nature's information about the actual alternative; usually $P_{i}\leq D_{i}$%
\cite{EnglertB}. For all pure states $\left| \Theta \right\rangle $, the
complementarity of single particle\cite{JaegerSV,Englert,EnglertB} 
\begin{equation}
V_{i}^{2}+D_{i}^{2}=1
\end{equation}
holds, while 
\begin{equation}
V_{i}^{2}+P_{i}^{2}\leq 1,
\end{equation}
where, by definition, the predictability $P_{i}$ of the path for particle 
{\it i} reads\cite{JaegerSV,Englert,EnglertB} 
\begin{equation}
P_{i}=\left| w_{i\uparrow }-w_{i\downarrow }\right| =\left| \left( \gamma
_{2}^{2}-\gamma _{3}^{2}\right) -\left( -1\right) ^{i}\left( \gamma
_{1}^{2}-\gamma _{4}^{2}\right) \right| 
\end{equation}
where $w_{ix}$ are  the probabilities that particle $i$ takes one way $%
\left| \uparrow \right\rangle $ or the other $\left| \downarrow
\right\rangle $. Comparing inequality (16) with Eq.(19), an equality 
\begin{equation}
V_{i}^{2}+V_{12}^{2}+P_{i}^{2}=1
\end{equation}
is achieved, which gives the relation among three physical quantities: $V_{i}
$, $V_{12}$ and $P_{i}$. In fact, Eq.(20) covers both of two complementarity
relations of inequalities (16) and (18) and implies the condition of
saturating the upper bound of the two inequalities. For example, if the
state of particle {\it i} is prepared symmetrically ($w_{i\uparrow
}=w_{i\downarrow }=\frac{1}{2}$) so that $P_{i}=0$, i.e., there is nothing
predictable about the paths in a symmetric one-particle counterpart, the
one- and two-particle complementarity will be of the form of an equality. Or
otherwise, if particle {\it i} is an asymmetric state ($w_{i\uparrow }\neq
w_{i\downarrow }$), that is, one path is more likely than the other to begin
with $P_{i}>0$, the complementarity is strictly limited in the inequality
(16), because partial WW information $P_{i}$ restrains full one-particle
interference visibility $V_{i}$.

As the complementarity is tightly associated with the property of the
composite system, it would be natural to examine the relationships between
these physical quantities and the entanglement $E$ of the system. By
calculating the von Neumann entropy\cite{Bennett}, the entanglement $E$ for
the pure state $\left| \Theta \right\rangle $ in Eq.(10) goes as 
\begin{equation}
E=-\frac{1-\sqrt{1-V_{12}^{2}}}{2}\log _{2}\left( \frac{1-\sqrt{1-V_{12}^{2}}%
}{2}\right) -\frac{1+\sqrt{1-V_{12}^{2}}}{2}\log _{2}\left( \frac{1+\sqrt{%
1-V_{12}^{2}}}{2}\right) .
\end{equation}
As illustrated in Fig. 2, this monotonically increasing relation between the
two-particle visibility $V_{12}$ and the entanglement $E$ indicates the
visibility of two-particle interference fringes $V_{12}$ is entirely
governed by the entanglement $E$ of pure two-particle source: the absence of
entanglement will results in zero-visibility two-particle interference
fringes (e.g., the $\left| \Phi \right\rangle $ state in Eq.(5), $E=0$, $%
V_{12}=0$), and the prefect entanglement, the full two-particle interference
fringes (e.g., the $\left| \Psi \right\rangle $ state in Eq.(4), $E=1$, $%
V_{12}=1$). From Eq.(21) one cannot directly make out clear relations
between $V_{i}$ or $D_{i}$ and $E$. However, from Eq.(20) and Eq.(17), $%
D_{i}^{2}=P_{i}^{2}+V_{12}^{2}$ is gained, which means that $D_{i}$ contains
both of the {\it a priori} WW information $P_{i}$ and the additional
information $V_{12}$ encoded in the entanglement between the object and the
WW maker due to the entire determination of $V_{12}$ by $E$. Since {\it a
priori} knowledge $P_{i}$ only lies on the feature of initial single
subsystem rather than that of the composite system, these quantities can be
studied with a given {\it a priori} WW knowledge $P_{i}$. For instance, in a
relevant symmetric one-particle interferometer $\left( P_{i}=0\right) $,
then $V_{i}=\sqrt{1-V_{12}^{2}}$ and $D_{i}=V_{12}$, which tie up with the
entanglement $E$ (see Fig.2): the entanglement $E$ is enhanced, more and
more Nature's WW information is stored in the states of the marker particle
through their correlation so that $D_{i}$ rises whereas the one-particle
visibility decreases, as discussed and experimentally testified in Ref. \cite
{Peng}. Most of one-particle interference experiments\cite
{Schwindt,Durr,Bertet,Peng} are discussed in this kind of symmetric
interferometer. If the interferometer begins with $P_{i}>0$, $V_{i}$ and $%
D_{i}$ still bear the similar dependence on the entanglement $E$, this
monotonically decreasing relation for $V_{i}$ and this monotonically
increasing relation for $D_{i}$ except for the wholly alterative amplitudes
based on {\it a priori} knowledge $P_{i}$, while $V_{12}$ does not alter the
relation on $E$ at all. Consequently, this figure also illuminates an
important view that quantum correlation is completely responsible for
complementarity.

In order to discuss these relations in greater detail, we study some special
families of states. When $\left| N\right| =\left| M\right| $, the following
two situations can be distinguished:

1) $\gamma _{1}\gamma _{4}=\gamma _{2}\gamma _{3}$: $\left| \Theta
\right\rangle $ corresponds to any product state ($E=0$), then $V_{12}=0$.
When $P_{i}=0$ (i.e., $\gamma _{1}=\gamma _{4}=\gamma _{2}=\gamma _{3}=\pm 
\frac{1}{2}$), we recover the $\left| \Phi \right\rangle $ state of Eq.(5)
(the sign makes no difference on these physical quantities), $V_{i}=1$. When 
$P_{i}>0$, the one-particle interference visibility is limited by $%
V_{i}^{2}+P_{i}^{2}=1$.

2) $\gamma _{2}=\gamma _{3}=0$ or $\gamma _{1}=\gamma _{4}=0$: $\left|
\Theta \right\rangle =\gamma _{1}\left| \uparrow \right\rangle _{1}\left|
\uparrow \right\rangle _{2}+\gamma _{4}\left| \downarrow \right\rangle
_{1}\left| \downarrow \right\rangle _{2}$ or $\left| \Theta \right\rangle
=\gamma _{2}\left| \uparrow \right\rangle _{1}\left| \downarrow
\right\rangle _{2}+\gamma _{3}\left| \downarrow \right\rangle _{1}\left|
\uparrow \right\rangle _{2}$, which are entangled. When $P_{i}=0$, (i.e., $%
\gamma _{1}=\gamma _{4}=\pm \frac{1}{\sqrt{2}}$ or $\gamma _{2}=\gamma
_{3}=\pm \frac{1}{\sqrt{2}}$), we retrieve the $\left| \Psi \right\rangle $
state of Eq.(4) or the similar state with $\left| \uparrow \right\rangle _{2}
$ and $\left| \downarrow \right\rangle _{2}$ interchanged which are the
maximal entangled states ($E=1$); in either case $V_{i}=0$ and $V_{12}=1$.
When $0<P_{i}<1$, ($\gamma _{1}\neq \gamma _{4}$ or $\gamma _{2}\neq \gamma
_{3}$), the $\left| \Theta \right\rangle $ state is partially entangled ($%
0<E<1$), it then follows $0<V_{12}<1$ which is limited by $%
V_{12}^{2}+P_{i}^{2}=1$ in the intermediate regime, but $V_{i}=0$ and $%
D_{i}=1$, the ''path'' can still be completely determined by a suitable
measurement. When $P_{i}=1$ (one of two $\gamma _{i}$ in $\left| \Theta
\right\rangle $ to be zero), the $\left| \Theta \right\rangle $ state again
recovers a product state ($E=0$), $V_{12}=0$ and $V_{i}=0$.

The cases discussed above are symmetric for particle 1 and particle 2 ($%
P_{1}=P_{2}$). We continue on investigating a special asymmetric case ($%
P_{1}\neq P_{2}$):

3) When $\left| \Theta \right\rangle $ is of this form 
\begin{equation}
\left| \psi \left( \alpha ,\beta \right) \right\rangle =\frac{1}{\sqrt{2}}%
\left[ \left| \uparrow \right\rangle _{1}\left( \cos \alpha \left| \uparrow
\right\rangle _{2}+\sin \alpha \left| \downarrow \right\rangle _{2}\right)
+\left| \downarrow \right\rangle _{1}\left( \cos \beta \left| \uparrow
\right\rangle _{2}+\sin \beta \left| \downarrow \right\rangle _{2}\right)
\right] ,
\end{equation}
In this case, $P_{1}=0$, $P_{2}=\left| \cos \left( \alpha +\beta \right)
\cos \left( \alpha -\beta \right) \right| $. Simple substitutions produce $%
V_{1}=\left| \cos \left( \alpha -\beta \right) \right| $, $V_{2}=\left| \sin
\left( \alpha +\beta \right) \cos \left( \alpha -\beta \right) \right| $ and 
$V_{12}=\left| \sin \left( \alpha -\beta \right) \right| $, which result in $%
V_{1}^{2}+V_{12}^{2}=1$ whereas $V_{2}^{2}+V_{12}^{2}\leq 1$. However,
whether for particle 1 or for particle 2, $V_{i}^{2}+V_{12}^{2}+P_{i}^{2}=1$
is satisfied. If $\alpha -\beta =\frac{\pi }{2}$, we again turn to the state
with $P_{1}=P_{2}=0$: 
\begin{equation}
\left| \psi \left( \alpha \right) \right\rangle =\frac{1}{\sqrt{2}}\cos
\alpha \left[ \left| \uparrow \right\rangle _{1}\left| \uparrow
\right\rangle _{2}+\left| \downarrow \right\rangle _{1}\left| \downarrow
\right\rangle _{2}\right] +\frac{1}{\sqrt{2}}\sin \alpha \left[ \left|
\uparrow \right\rangle _{1}\left| \downarrow \right\rangle _{2}+\left|
\downarrow \right\rangle _{1}\left| \uparrow \right\rangle _{2}\right] .
\end{equation}
which was discussed in Ref.\cite{JaegerHS}: $V_{i}=\left| \sin \left(
2\alpha \right) \right| $ and $V_{12}=\left| \cos \left( 2\alpha \right)
\right| $ which make $V_{i}^{2}+V_{12}^{2}=1$.

\section{Experimental investigations}

In our experiments, the chosen quantum system is $^{13}$C-labeled chloroform 
$^{13}$CHCl$_{3}$ (Cambridge Isotope Laboratories, Inc.) molecules with the
hydrogen nuclei ($^{1}$H) for particle 1 and the carbon nuclei ($^{13}$C)
for particle 2. The spin-spin coupling constant $J$ between $^{13}$C and $%
^{1}$H is 214.95 Hz. The relaxation times were measured to be $T_{1}=4.8$ $%
sec$ and $T_{2}=3.3$ $sec$ for the proton, and $T_{1}=17.2$ $sec$ and $%
T_{2}=0.35$ $sec$ for carbon nuclei. Experiments were performed on a
BrukerARX500 spectrometer with a probe tuned at 125.77MHz for $^{13}$C and
at 500.13MHz for $^{1}$H by using a conventional liquid-state NMR techniques.

The quantum ensemble is firstly prepared in a pseudo-pure state $|\psi
_{0}\rangle =\left| \uparrow \right\rangle _{1}\left| \uparrow \right\rangle
_{2}$ from thermal equilibrium by applying line-selective pulses with
appropriate frequencies and rotation angles as well as a consequent magnetic
gradient pulse\cite{Peng1}. Different two-particle source which corresponds
to any $\left| \Theta \right\rangle $ state in Eq.(10) can be prepared from
the $|\psi _{0}\rangle $ state by appropriate operatations, e.g., the
entangled state $\left| \Psi \right\rangle $ of Eq.(4), the product state $%
\left| \Phi \right\rangle $ of Eq (5) and the special family of states $%
\left| \psi \left( \alpha \right) \right\rangle $ of Eq.(23) by applying,
respectively, the following NMR pulse sequences: $Y_{1}(-\frac{\pi }{2}%
)X_{1}(-\frac{\pi }{2})Y_{1}(\frac{\pi }{2})X_{2}(-\frac{\pi }{2}%
)Y_{2}\left( \frac{\pi }{2}\right) J_{12}(\frac{\pi }{2})Y_{2}(\frac{\pi }{2}%
),$ $Y_{1}(\frac{\pi }{2})Y_{2}(\frac{\pi }{2})$ and $Y_{2}\left( \frac{\pi 
}{2}\right) X_{2}(\frac{\pi }{2})J_{12}(\frac{\pi }{2}-2\alpha )X_{2}\left( -%
\frac{\pi }{2}\right) Y_{1}(\frac{\pi }{2})$ to be read from left to right.
Here $Y_{i}(\theta )$ denotes an $\theta $ rotation about $\hat{y}$ axis on
particle $i$ and so forth, and $J_{12}(\varphi )$ represents a time
evolution of $\varphi /\pi J_{12}$ under the scalar coupling between spins 1
and 2. Then the transformation $U_{i}\left( \phi _{i}\right) $ to achieve
the operations of phase shifters and beam splitters was realized by the NMR
pulse sequence $X_{i}\left( -\theta _{1}\right) Y_{i}\left( \theta
_{2}\right) X_{i}\left( -\theta _{1}\right) $ with $\theta _{1}=\tan
^{-1}(-\sin \phi _{i})$ and $\theta _{2}=2sin^{-1}(-cos\phi _{i}/\sqrt{2})$. 
$U_{i}\left( \phi _{i}\right) $ with a set of appropriate values $\theta _{1}
$ and $\theta _{2}$ was repeatedly applied in experiments to vary the value
of $\phi _{i}$ from $0$ to $2\pi $. Finally, the probabilities of single and
joint measurements were obtained by quantum state tomography\cite{Chuang} to
reconstruct the diagonal elements of the output density matrices, which can
be completed by employing reading-out pulses $Y_{1}(\frac{\pi }{2})$ and $%
Y_{2}(\frac{\pi }{2})$ after a gradient pulse field to record $^{1}$H and $%
^{13}$C spectra, respectively. The quantitative measurements of the one- and
two-particle interference visibilities were gained by monitoring the
variations of these probabilities versus $\phi _{i}$. In our experiments,
the complementarity test is restricted in the cases of $P_{i}=0$, that is,
to verify $V_{i}^{2}+V_{12}^{2}=1$.

For two extreme cases, we simultaneously varied the values of $\phi _{1}$
and $\phi _{2}$ with the respective increment of $\pi /16$. Applying the
procedure stated above, the experimental results are shown in Fig. 3. As
expected, for the entangled state $\left| \Psi \right\rangle $, two-particle
interference fringes are displayed (Fig. 3(b)) but almost no one-particle
interference fringes (Fig. 3(a)). and the opposite situation for the product
state $\left| \Phi \right\rangle $ (Fig. 3(c) and (d)). From these
experimental data points in Fig. 3, the measured values of the visibilities
were extracted: $V_{1}=0.12,$ $V_{2}=0.14$, $V_{12}=0.87$ for the entangled
state $\left| \Psi \right\rangle $ and $V_{1}=0.93,$ $V_{2}=0.99$, $%
V_{12}=0.10$ for the product state $\left| \Phi \right\rangle $. By contrast
with the theoretical expectations (see Sec II), more experimental errors
were introduced in the case of $\left| \Psi \right\rangle $ inasmuch as the
preparation of the entangled state $\left| \Psi \right\rangle $ was more
complicated than that of the product state $\left| \Phi \right\rangle $.

The interferometric complementarity of the special family of states $\left|
\Psi \left( \alpha \right) \right\rangle $ in the intermediate regime was
testified with a similar procedure. As $\overline{p}\left( K_{1}K_{2}\right) 
$ is a function of $\phi _{1}$ and $\phi _{2}$ whose extrema $\overline{p}%
_{\max ,\min }\left( K_{1}K_{2}\right) $ were found at $(\phi _{1},\phi
_{2})=\left( n\pi +\frac{\pi }{2},m\pi +\frac{\pi }{2}\right) $ in the
theoretical calculations, we scanned one of $\phi _{i}$ while fixing the
other into $\pi /2$ for a given state $\left| \Psi \left( \alpha \right)
\right\rangle $, then repeated experiment with interchanging them, instead
of simultaneously scanning $\phi _{1}$ and $\phi _{2}$. The procedure was
repeated for different $\alpha $. For the convenience of experiments, we
changed the values of $\alpha $ from $\pi /4$ to $21\pi /16$ with the
increment of $\pi /16$. From the desirable interference fringes shown from
variations of the normalized populations versus $\phi _{i}$, the
visibilities were obtained. The measured and theoretical visibilities of
one- and two-particle interference $V_{1}\left( \alpha \right) ,$ $%
V_{2}\left( \alpha \right) $ and $V_{12}\left( \alpha \right) $ are plotted
in Fig. 4(a), together with entanglement $E\left( \alpha \right) $ of the
two-particle system. These experimental results are in good agreement with
the theoretical expectations. This figure also shows the relationship
between the visibilities and entanglement, i.e., $E\left( \alpha \right) $
varies synchronously with $V_{12}\left( \alpha \right) $ and has the
opposite variation trend with $V_{i}\left( \alpha \right) $, consistent with
the theoretical analyses in Sec.III. The experimental data of the
complementarity relations for $V_{i}^{2}\left( \alpha \right)
+V_{12}^{2}\left( \alpha \right) $ are also depicted in Fig. 4(b). The
initial state of two-particle source is a symmetric state with no {\it a
priori} WW knowledge of particle 1 and 2 ($P_{1}=P_{2}=0$), so that all data
in Fig. 4(b) should be unity.

In these experiments, the estimated errors are less than $\pm 10\%$ due to
the inhomogeneity of the RF field and static magnetic field, imperfect
calibration of RF pulses and signal decay during the experiments. If we take
into account the imperfections of the experiments, the measured data in our
NMR experiments agree well with theory.

\section{Conclusion}

For a general pure two-particle source, we derive the interferometric
complementarity between one- and two-particle interference visibilities and
obtain an equality among the one-particle interference visibility $V_{i}$,
two-particle interference visibility $V_{12}$ and the predication $P_{i}$ of
the path of single particle: $V_{i}^{2}+V_{12}^{2}+P_{i}^{2}=1$. The
equality not only embodies two complementarity relations of $%
V_{i}^{2}+V_{12}^{2}\leq 1$ and $V_{i}^{2}+P_{i}^{2}\leq 1$, but also
implies the conditions of their saturating the upper bound of them, e.g.,
for the one- and two-particle interference, the duality of $%
V_{i}^{2}+V_{12}^{2}=1$ holds only when the interferometer involves a
symmetric single-particle counterpart (i.e., $P_{i}=0$). In fact, the
equality is also a demenstration of the relationships among one-paritcle
wave-like ($V_{i}$), particle-like ($P_{i}$) and two-particle wave-like ($%
V_{12}$) attributes. In addition, we also manifest the role of entanglement
in this type of interference experiments, which reveals whether in a
symmetric way or in an asymmetric one, the greater entanglement is the
quantum composite system, the lower the visibility of two-particle
interference fringes whereas the higher the visibility of two-particle
interference fringes. Meanwhile, we also consider the relation between two
interferometric complementarities. The one- and two-particle interference
complementarity has been testified for a kind of special symmetric
interferometer ($P_{1}=P_{2}=0$) in a spin NMR ensemble, including two
extreme cases and a special family of states. The experimental results
showed the theoretical predictions of quantum mechanics within the
experimental errors. Though the test was performed over interatomic
distances, the experimental scheme originates from quantum version and their
dynamical evolution is quantum mechanical. Therefore, these experiments also
indicate that non-classical properties associated with entanglement can be
displayed in NMR ensemble as long as the initial state is well prepared.

\begin{center}
{\bf ACKNOWLEDGMENTS}
\end{center}

This work is supported by the National Natural Science Foundation of China
(Grant NO. 10274093) and the National Fundamental Research Program
(2001CB309300). We also thank Hanzheng Yuan, Zhi Ren, and Xu Zhang for help
in the course of experiments.

\begin{center}
{\large Figure Captions}
\end{center}

Fig. 1 Schematic two-particle interferometer using beam splitters BS$_{1}$,
BS$_{2}$ and phase shifters $\phi _{1}$, $\phi _{2}$.

Fig. 2 The entanglement $E$ versus visibility $V_{12}$ for pure two-particle
source (denoted by the solid line), visibility $V_{i}$ and
distinguishability $D_{i}$ of a single particle in different {\it a priori}
WW knowledge $P_{i}$: the dotted line of $V_{i}$ and the solid line of $%
D_{i}=V_{12}$ for $P_{i}=0$; the dashed line of $V_{i}$ and the dashdotted
line of $D_{i}=V_{12}$ for $P_{i}=0.4$.

Fig. 3 The single and ``corrected'' joint probabilities detected in the one-
and two-particle interference, (a) and (b) for the entangled state $\left|
\Psi \right\rangle $, (c) and (d) for the product state $\left| \Phi
\right\rangle $. In (a) and (c), data points $\square ,$ $+,$ $\bigcirc $
and $\times $ denote the single probabilities $p\left( \left| \uparrow
\right\rangle _{1}\right) ,p\left( \left| \downarrow \right\rangle
_{1}\right) ,p\left( \left| \uparrow \right\rangle _{2}\right) $ and $%
p\left( \left| \downarrow \right\rangle _{2}\right) $, respectively; in (b)
and (d), $\bigcirc ,$ $\times ,$ $\square $ and $+$ denote the ``corrected''
joint probabilities $\overline{p}\left( \left| \uparrow \right\rangle
_{1}\left| \uparrow \right\rangle _{2}\right) ,$ $\overline{p}\left( \left|
\uparrow \right\rangle _{1}\left| \downarrow \right\rangle _{2}\right) ,$ $%
\overline{p}\left( \left| \downarrow \right\rangle _{1}\left| \uparrow
\right\rangle _{2}\right) $ and $\overline{p}\left( \left| \downarrow
\right\rangle _{1}\left| \downarrow \right\rangle _{2}\right) $,
respectively. Theoretical curves are depicted with the solid lines.

Fig. 4 The visibilities and the complementarity relation of a special family
of states $\left| \psi \left( \alpha \right) \right\rangle $ as a function
of the angle $\alpha .$ (a) Data points $\bigcirc ,$ $+$ and $*$ denote,
respectively, the visibilities of one-particle interference $V_{1},$ $V_{2}$
and two-particle interference $V_{12}$. The solid lines are the theoretical
expectations and the dotted line denotes the theoretical curve of
entanglement $E$. (b) Experimental test of the complementarity relation $%
V_{1}^{2}+V_{12}^{2}$ (denoted by data ponits $\square $) and $%
V_{2}^{2}+V_{12}^{2}$ (denoted by data ponits $\triangledown $) are plotted
as a function of $\alpha $. The solid line represents the theoretical
expectations.


\begin{references}
\bibitem{Bohr}  N. Bohr, 1928 Naturwissenschaften {\bf 16} 245; Nature
(London) {\bf 121} 580 (1928).

\bibitem{Feynman}  R. P. Feynman, R. B. Leifhton, and M. Sands, the Feynamn
Lectures of Physics, Vol. III. Quantum Mechanics, Addison -Wesley, Reading
(1965).

\bibitem{Scully}  M. O. Scully, B. -G. Englert and H. Walther, 1991 Nature 
{\bf 351} 111.

\bibitem{Wootters}  W. K. Wootters and W. H. Zurek, Phys. Rev. D {\bf 19}
473 (1979).

\bibitem{Bartell}  L. S. Bartell, Phys. Rev. D {\bf 21} 1698 (1980).

\bibitem{Greenberger}  D. M. Greenberger and A. Yasin, Phys. Lett. A {\bf 128%
} 391 (1988).

\bibitem{Mandel}  L. Mandel, Opt. Lett. {\bf 16} 1882 (1991).

\bibitem{JaegerSV}  G. Jaeger, A. Shimony and L. Vaidman, Phys. Rev. A {\bf %
51} 54 (1995).

\bibitem{Englert}  B. -G. Englert, Phys. Rev. Lett. {\bf 77,} 2154 (1996).

\bibitem{EnglertB}  B. -G. Englert, and J. A. Bergou, Opt. Comm. 179, 337
(2000).

\bibitem{Taylor}  G. I. Taylor, Proc. Camb. Phil. Soc. {\bf 15,} 114 (1909).

\bibitem{Mittelstaedt}  P. Mittelstaedt, A. Prieur and R. Schieder, Found.
Phys. {\bf 17,} 891 (1987).

\bibitem{Schwindt}  P. D. D. Schwindt, P. G. Kwiat and B. -G. Englert, Phys.
Rev. A {\bf 60,} 4285 (1999).

\bibitem{Mollenstedt}  G. M$\ddot{o}$llenstedt and C. J$\ddot{o}$nsson, Z.
Phys. {\bf 155,} 472 (1959); A. Tonomura, J. Endo, T. Matsuda, and T.
Kawasaki, Am. J. Phys. {\bf 57,} 117 (1989).

\bibitem{Zeilinger}  A. Zeilinger, R. G$\ddot{a}$hler, C. G. Shull, W.
Treimer, and W. Mampe, Rev. Mod. Phys. {\bf 60,} 1067 (1988).

\bibitem{Rauch}  H. Rauch and J. Summhammer, Phys. Lett. A {\bf 104,} 44
(1984).

\bibitem{Summhammer}  J. Summhammer, H. Rauch and D. Tuppinger, Phys. Rev. A 
{\bf 36,} 4447 (1987).

\bibitem{Carnal}  O. Carnal and J. Mlynek, Phys. Rev. Lett. {\bf 66,} 2689
(1991).

\bibitem{Durr}  S. D$\ddot{u}$rr, T. Nonn and G. Rempe, Phys. Rev. Lett. 
{\bf 81,} 5705 (1998).

\bibitem{Bertet}  P. Bertet, S. Osnaghl, A. Rauschenbeutel, G. Nogues, A.
Auffeves, M. Brune, J. M. Ralmond and S. Haroche, Nature {\bf 411,} 166
(2001).

\bibitem{Zhu}  X. Zhu, X. Fang, X. Peng, M. Feng, K. Gao and F. Du, J. Phys.
B {\bf 34,} 4349 (2001).

\bibitem{Peng}  X. Peng, X. Zhu, X. Fang, M. Feng, M. Liu, and K. Gao, J.
Phys. A 36, 2555 (2003).

\bibitem{Ghosh}  R. Ghosh and L. Mandel, Phys. Rev. Lett. {\bf 59}, 1903
(1987).

\bibitem{All}  C. O. Alley and Y. H. Shih, in {\it Proceedings of the Second
International Symposium on Foundations of Quantum Mechanics in Light of New
Technology}, edited by M. Namiki {\it et al}. (Physical Society of Japan,
Tokyo, 1986), p. 47; Y. H. Shih and C. O. Alley, Phys. Rev. Lett. {\bf 61},
2921 (1988); C. K. Hong, Z. Y. Ou, and L. Mandel, Phys. Rev. Lett. {\bf 59},
2044 (1987); J. D. Franson, {\it ibid}. {\bf 64}, 2495 (1990); P. G. Kwiat,
W. A. Vereka, C. K. Hong, H. Nathel, and R. Y. Chiao, Phys. Rev. A {\bf 41},
2910 (1990); M. Horne, A. Shimony, and A. Zeilinger, in {\it Quantum
Coherence}, edited by J. Anandan (World Scientific, Singapore, 1990), p. 356.

\bibitem{HorneZeil}  M. A. Horne and A. Zeilinger in {\it Proceedings of the
Symposium on the Foundations of Modern Physics}, edited by P. Lahti and P.
Mittelstaedt (World Scientific, Singapore, 1985), p435.

\bibitem{JaegerHS}  G. Jaeger, M. A. Horne and A. Shimony, Phys. Rev. A {\bf %
48}, 1023 (1993).

\bibitem{Abouraddy}  A. F. Abouraddy, M. B. Nasr, B. E. A. Saleh, A. V.
Sergienko and M. C. Teich, Phys. Rev. A {\bf 63}, 063803 (2001).

\bibitem{Bennett}  C. H. Bennett,H. J. Bernstein S. Popescu B. Schumacher,
Phys. Rev. A {\bf 53} 2046 (1996).

\bibitem{Peng1}  X. Peng, X. Zhu, X. Fang, M. Feng, K. Gao, X. Yang and M.
Liu, Chem. Phys. Lett. {\bf 340} 509 (2001).

\bibitem{Chuang}  I. L. Chuang, N. Gershenfeld, M. Kubinec and D. Leung,
Proc. Roy. Soc. Lond A {\bf 454,} 447 (1998).\newpage
\end{references}
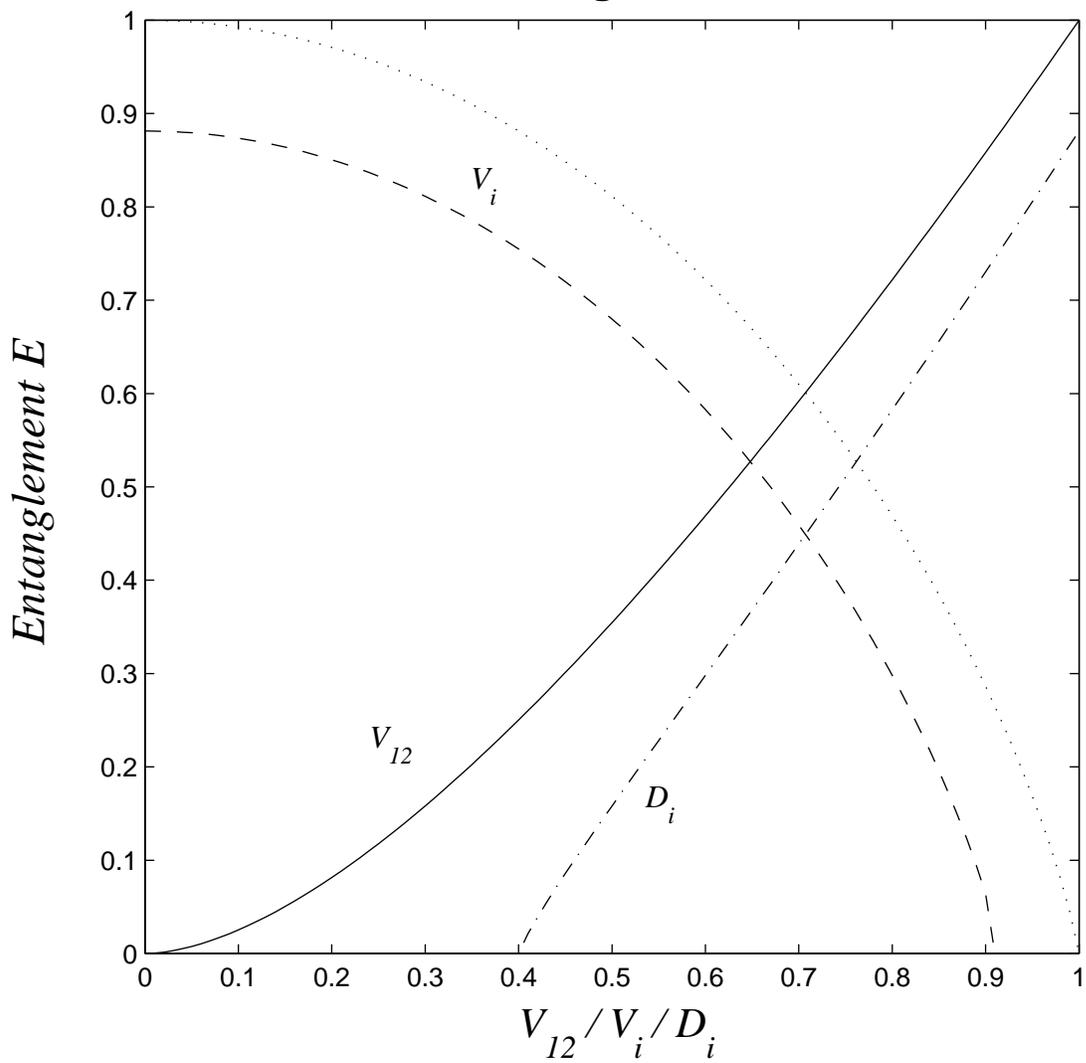
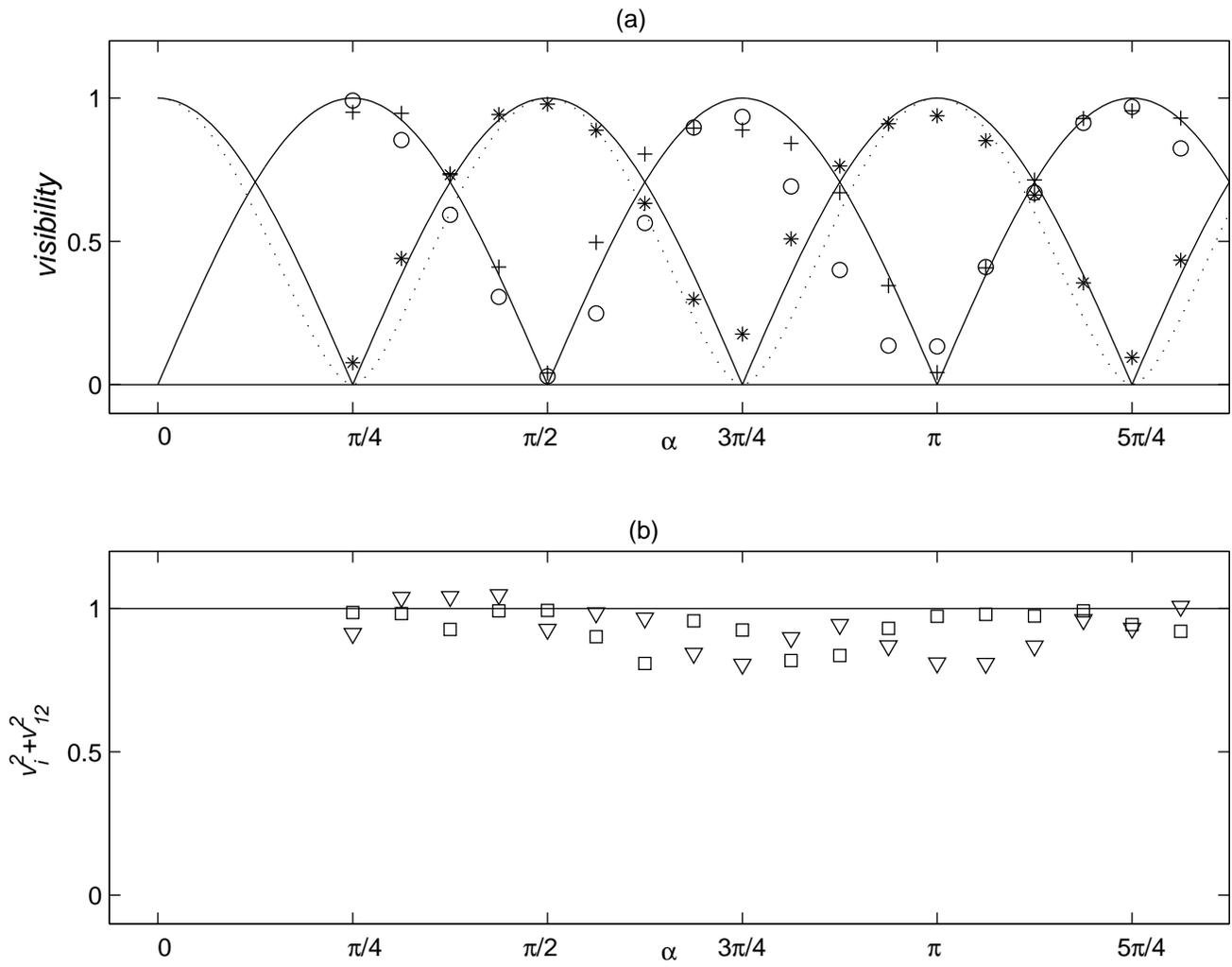
\end{document}